\documentstyle[12pt]{article}

\catcode`\@=11
\long\def\@makefntext#1{
\protect\noindent \hbox to 3.2pt {\hskip-.9pt
$^{{\ninerm\@thefnmark}}$\hfil}#1\hfill}		

\def\@makefnmark{\hbox to 0pt{$^{\@thefnmark}$\hss}}  

\def\ps@myheadings{\let\@mkboth\@gobbletwo
\def\@oddhead{\hbox{}
\rightmark\hfil\ninerm\thepage}
\def\@oddfoot{}\def\@evenhead{\ninerm\thepage\hfil
\leftmark\hbox{}}\def\@evenfoot{}
\def\sectionmark##1{}\def\subsectionmark##1{}}

\renewcommand{\thefootnote}{\fnsymbol{footnote}}

\newcounter{sectionc}\newcounter{subsectionc}\newcounter{subsubsectionc}
\renewcommand{\section}[1] {\vspace*{0.6cm}\addtocounter{sectionc}{1}
\setcounter{subsectionc}{0}\setcounter{subsubsectionc}{0}\noindent
	{\normalsize\bf\thesectionc. #1}\par\vspace*{0.4cm}}
\renewcommand{\subsection}[1] {\vspace*{0.6cm}\addtocounter{subsectionc}{1}
	\setcounter{subsubsectionc}{0}\noindent
	{\normalsize\it\thesectionc.\thesubsectionc. #1}\par\vspace*{0.4cm}}
\renewcommand{\subsubsection}[1]
{\vspace*{0.6cm}\addtocounter{subsubsectionc}{1}
	\noindent {\normalsize\rm\thesectionc.\thesubsectionc.\thesubsubsectionc.
	#1}\par\vspace*{0.4cm}}

\newcounter{appendixc}
\newcounter{subappendixc}[appendixc]
\newcounter{subsubappendixc}[subappendixc]

\renewcommand{\appendix}[1] {\vspace*{0.6cm}
        \refstepcounter{appendixc}
        \setcounter{figure}{0}
        \setcounter{table}{0}
        \setcounter{equation}{0}
        \renewcommand{\thefigure}{\Alph{appendixc}.\arabic{figure}}
        \renewcommand{\thetable}{\Alph{appendixc}.\arabic{table}}
        \renewcommand{\theappendixc}{\Alph{appendixc}}
        \renewcommand{\theequation}{\Alph{appendixc}.\arabic{equation}}
        \noindent{\bf Appendix \theappendixc #1}\par\vspace*{0.4cm}}

\def\abstracts#1{{

\centering{\begin{minipage}{12.2truecm}\footnotesize\baselineskip=12pt\noindent
	\centerline{\footnotesize ABSTRACT}\vspace*{0.3cm}
	\parindent=0pt #1
	\end{minipage}}\par}}

\renewenvironment{thebibliography}[1]
	{\begin{list}{\arabic{enumi}.}
	{\usecounter{enumi}\setlength{\parsep}{0pt}
\setlength{\leftmargin 1.25cm}{\rightmargin 0pt}
	 \setlength{\itemsep}{0pt} \settowidth
	{\labelwidth}{#1.}\sloppy}}{\end{list}}

\newcounter{itemlistc}
\newcounter{romanlistc}
\newcounter{alphlistc}
\newcounter{arabiclistc}

\newcommand{\fcaption}[1]{
        \refstepcounter{figure}
        \setbox\@tempboxa = \hbox{\footnotesize Fig.~\thefigure. #1}
        \ifdim \wd\@tempboxa > 6in
           {\begin{center}
        \parbox{6in}{\footnotesize\baselineskip=12pt Fig.~\thefigure. #1}
            \end{center}}
        \else
             {\begin{center}
             {\footnotesize Fig.~\thefigure. #1}
              \end{center}}
        \fi}

\newcommand{\tcaption}[1]{
        \refstepcounter{table}
        \setbox\@tempboxa = \hbox{\footnotesize Table~\thetable. #1}
        \ifdim \wd\@tempboxa > 6in
           {\begin{center}
        \parbox{6in}{\footnotesize\baselineskip=12pt Table~\thetable. #1}
            \end{center}}
        \else
             {\begin{center}
             {\footnotesize Table~\thetable. #1}
              \end{center}}
        \fi}

\def\@citex[#1]#2{\if@filesw\immediate\write\@auxout
	{\string\citation{#2}}\fi
\def\@citea{}\@cite{\@for\@citeb:=#2\do
	{\@citea\def\@citea{,}\@ifundefined
	{b@\@citeb}{{\bf ?}\@warning
	{Citation `\@citeb' on page \thepage \space undefined}}
	{\csname b@\@citeb\endcsname}}}{#1}}

\newif\if@cghi
\def\cite{\@cghitrue\@ifnextchar [{\@tempswatrue
	\@citex}{\@tempswafalse\@citex[]}}
\def\citelow{\@cghifalse\@ifnextchar [{\@tempswatrue
	\@citex}{\@tempswafalse\@citex[]}}
\def\@cite#1#2{{$\null^{#1}$\if@tempswa\typeout
	{IJCGA warning: optional citation argument
	ignored: `#2'} \fi}}

 1
 1
 1

\font\ninerm=cmr9

\def\h{\textstyle{1\over 2}}

\begin{document}

\newcommand{\st}{\scriptstyle}
\newcommand{\sst}{\scriptscriptstyle}
\newcommand{\mco}{\multicolumn}
\newcommand{\epp}{\epsilon^{\prime}}
\newcommand{\vep}{\varepsilon}
\newcommand{\ra}{\rightarrow}
\newcommand{\ppg}{\pi^+\pi^-\gamma}
\newcommand{\vp}{{\bf p}}
\newcommand{\ko}{K^0}
\newcommand{\kb}{\bar{K^0}}
\newcommand{\al}{\alpha}
\newcommand{\ab}{\bar{\alpha}}
\def\be{\begin{equation}}
\def\ee{\end{equation}}
\def\bea{\begin{eqnarray}}
\def\eea{\end{eqnarray}}
\def\CPbar{\hbox{{\rm CP}\hskip-1.80em{/}}}

\hbox{  }
\vspace{-1.1in}
\noindent
\hfill CTP-TAMU-04/95

\noindent
\hfill ACT-01/95

\centerline{\normalsize\bf EFFECTIVE SUPERGRAVITY FROM 4-D FERMIONIC STRINGS
\footnote{To appear in the proceedings of ``Beyond the Standard Model IV'',
December 13--18, 1994, Lake Tahoe, CA.}}
\baselineskip=22pt

\centerline{\footnotesize KAJIA YUAN}
\baselineskip=13pt
\centerline{\footnotesize\it Center for Theoretical Physics,
Department of Physics, Texas A \& M University}
\baselineskip=12pt
\centerline{\footnotesize\it College Station, TX 77843-4242, USA}
\baselineskip=13pt
\centerline{\footnotesize\it Astroparticle Physics Group,
Houston Advanced Research Center (HARC)}
\baselineskip=12pt
\centerline{\footnotesize\it The Woodlands, TX 77381, USA}
\centerline{\footnotesize E-mail: kyuan@tac.harc.edu}

\vspace*{0.4cm}
\abstracts{The untwisted moduli in four-dimensional fermionic string
models are identified by considering world-sheet Abelian Thirring
interactions. The full K\"ahler potentials of untwisted sector fields
are given for a restricted class of models.}

\vspace*{0.35cm}
\normalsize\baselineskip=15pt
\setcounter{footnote}{0}
\renewcommand{\thefootnote}{\alph{footnote}}

String-derived effective supergravity theories have two salient
features which made them of great theoretical and phenomenological
interest: (1) the three standard functions $(f_{ab}, K, W)$
that specify the $N=1$ supergravity Lagrangian are calculable;
and (2) the underlying conformal invariance entails the existence
of moduli. In this talk, I shall give a brief account of
our recent work,\cite{LNY} which concerns the effective supergravity
theories from a class of four-dimensional fermionic string models.\cite{FFF}

Let me first describe a procedure with which the
untwisted moduli in fermionic models can be identified. In the context
of conformal field theory (CFT), moduli correspond to exactly marginal
operators which generate deformations of a CFT that preserve conformal
invariance at the classical as well as quantum level. Therefore,
for a given free fermionic model, to see whether or not a particular
massless scalar field is a modulus field, one would first construct
its zero-momentum vertex operator in the zero-ghost-charge picture
and then examine if this operator is exactly marginal or not. For the
massless scalar states form the Neveu-Schwarz sector, the corresponding
vertex operators are given by world-sheet Thirring interactions.
A subset of them, {\it ie.}, the {\it Abelian} Thirring
operators of the form $J^i_L(z){\bar J}^j_R({\bar z})$, where
$J^i_L, {\bar J}^j_R$ are some $U(1)$ chiral currents realized by
world-sheet fermions, are exactly marginal. One way to show this
statement is to perform a bosonization, upon which the operators
$J^i_L(z){\bar J}^j_R({\bar z})$ become $\partial X^I{\bar \partial}X^J$,
the standard form of the exactly marginal operators for the untwisted moduli
in orbifold models. In addition, it is obvious that such Abelian Thirring
operators satisfy the necessary and sufficient condition for
integrability.\cite{CS}

In practice, many fermionic models contain an $N=4$ submodel with an
$SO(44)$ gauge group arising from the right-movers and an $U(1)^6$
from the left-movers. At this $N=4$ level, the untwisted massless scalar
states transform in the adjoint representation of $SO(44)$, and the
$6\times 22$ states in the Cartan subalgebra provide the moduli which
parameterize the coset space $SO(6,22)/SO(6)\times SO(22)$. In the
original models, not all of these states are compatible with the
spin-structures, and only those of them which survive all the GSO-projections
give rise to the compatible exactly marginal operators and thus
are the untwisted moduli.

As an example, consider a model with basis
${\cal B}=\{{\bf 1},S,b_1,b_2,b_3\}$, where
\begin{eqnarray}
S&=(1\ 100\ 100\ 100\ 100\ 100\ 100\ :\
\overbrace{000000}^{\textstyle\bar y^I}\
\overbrace{000000}^{\textstyle\bar \omega^I}\
\!\!\overbrace{00000\ 000\ 0_8}^{\textstyle\bar\Psi^{\pm a}}),\\
b_1&=(1\ 100\ 100\ 010\ 010\ 010\ 010\ :\ 001111\ 000000\ 11111\ 100\
0_8),\\
b_2&=(1\ 010\ 010\ 100\ 100\ 001\ 001\ :\ 110000\ 000011\ 11111\ 010\
0_8),\\
b_3&=(1\ 001\ 001\ 001\ 001\ 100\ 100\ :\ 000000\ 111100\ 11111\ 001\
0_8).
\end{eqnarray}
By tracking the allowed Abelian Thirring terms, one can explicitly find
the untwisted moduli fields of this model, which parameterize the following
moduli space
\begin{equation}
{\cal M}={SO(2,2)\over {SO(2)\times SO(2)}}\otimes
{SO(2,2)\over {SO(2)\times SO(2)}}\otimes
{SO(2,2)\over {SO(2)\times SO(2)}}.
\end{equation}
This result is in complete agreement with the fact that this simple
fermionic model can be viewed as a symmetric $Z_2\times Z_2$ orbifold
model.

In several realistic fermionic models, there are spin-structure
vectors which assign {\it asymmetrically} the boundary conditions for
the left-moving real fermions $y^I, \omega^I$ relative to the
right-moving ones ${\bar y}^I,{\bar\omega}^I$, such models can be
interpreted as asymmetric orbifold models. For instance, in the
``revamped'' flipped $SU(5)$ model,\cite{FLI} the basis vector
responsible for the ``asymmetry'' is given by
\begin{equation}
\alpha=(0\ 000\ 000\ 000\ 011\ 000\ 011\ :\ 000101\ 011101\
\h\h\h\h\h\ \h\h\h\ \h\h\h\h\ 1100).
\end{equation}
And the untwisted moduli space of this model reduces to
\begin{equation}
{\cal M}={SO(2,1)\over SO(2)}\otimes
{SO(2,1)\over SO(2)}\otimes
{SO(2,2)\over {SO(2)\times SO(2)}}.
\end{equation}

I now discuss the K\"ahler potential for this class of fermionic models.
As can be seen from above examples, the untwisted moduli fields split up
into three sets, the fields in each set parameterize either
$SO(2,2)/SO(2)\times SO(2)$ or $SO(2,1)/SO(2)$. Similarly,
the untwisted matter fields, namely, those states which correspond
to the non-zero roots of $D_{22}$ at the $N=4$ level, also fall into
three sets. Using $S$-matrix approach, by computing various four-point
string scattering amplitudes, it is found that the scalars in each set
admit a non-linear $\sigma$-model structure of
$SO(2,n)/SO(2)\times SO(n)$, where $n$ counts the total number of
scalar fields in the set, including both the moduli and the matter
fields. In addition, in terms of the actual massless string states,
the most natural form of the K\"ahler potential for each set is
\begin{equation}
K(\alpha_i,{\bar\alpha}_i)=-\log\left(1-\sum_i^n\alpha_i{\bar\alpha}_i
+{1\over 4}\left|\sum_i^n\alpha^2_i\right|^2\right).
\label{eq:kkk}
\end{equation}
A different approach to this problem for some simple models of the
same class was put forward sometime ago.\cite{FGKP} However,
the $S$-matrix analysis enables one to establish the precise relation
between the complex coordinates $\alpha_i$ in (\ref{eq:kkk}) and the
massless string states.\cite{LNY}

One can recast the K\"ahler potential of form (\ref{eq:kkk}) into
the so-called ``supergravity basis'', in which the phenomenological
studies as well as the theoretical analysis of some problems can
be carried out in the usual fashion. To this end, in addition to
replace the the ``string basis'' moduli fields by the corresponding
fields $T,U$, some holomorphic field redefinitions for the matter fields
are also necessary.\cite{LNY} For the case of moduli space
$SO(2,2)/SO(2)\times SO(2)$, up to a K\"ahler transformation, one has
\begin{eqnarray}
K&=&-\log\left\{(T+{\bar T})(U+{\bar U})
-\sum^n_iA_i{\bar A}_i
+{1\over 4}{(T_c+{\bar T}_c)\over |{\bar T}_c+T|^2}
{(U_c+{\bar U}_c)\over |{\bar U}_c+U|^2}\left|\sum^n_iA^2_i\right|^2
\right. \nonumber\\
&&\left.+{1\over 2}\left({{T_c-T}\over {T_c+{\bar T}}}\right)
\left({{U_c-U}\over {U_c+{\bar U}}}\right)\sum^n_i{\bar A}^2_i
+{1\over 2}\left({{{\bar T}_c-{\bar T}}\over {{\bar T}_c+T}}\right)
\left({{{\bar U}_c-{\bar U}}\over {{\bar U}_c+U}}\right)\sum^n_iA^2_i\right\};
\end{eqnarray}
while for the case of moduli space $SO(2,1)/SO(2)$, one has
\begin{eqnarray}
K&=&-\log\left\{(T+{\bar T})-\sum^n_iA_i{\bar A}_i
+{1\over 4}{(T_c+{\bar T}_c)\over |{\bar T}_c+T|^2}\left|\sum^n_iA^2_i\right|^2
\right. \nonumber\\
&&\left.+{1\over 2}
{1\over {|{\bar T}_c+T|+\sqrt{(T+{\bar T})(T_c+{\bar T}_c)}}}
\left[{(T_c-T)^2\over |{\bar T}_c+T|}\sum^n_i{\bar A}^2_i+h.c.\right]\right\}.
\end{eqnarray}
It can be shown that these K\"ahler potentials are invariant under target
space duality transformations.\cite{LNY}

In conclusion, I have illustrated how to identify untwisted moduli in
fermionic string models, and presented the K\"ahler potentials for a
restricted class of models.

\vspace*{0.5cm}
\noindent {\normalsize\bf Acknowledgements}

This work has been supported by the ICSC-World Laboratory.


\vspace*{0.5cm}
\noindent {\normalsize\bf References}

\begin{thebibliography}{9}
\bibitem{LNY} J. Lopez, D. V. Nanopoulos and K. Yuan, {\it Phys. Rev.}
{\bf D50} (1994) 4060.
\bibitem{FFF} H. Kawai, D. C. Lewellen and S.-H. Tye, {\it Nucl. Phys.}
{\bf B288} (1987) 1; I. Antoniadis, C. Bachas and C. Kounnas, {\it Nucl. Phys.}
{\bf B289} (1987) 87; I. Antoniadis and C. Bachas, {\it Nucl. Phys.}
{\bf B298} (1988) 586.
\bibitem{CS} S. Chaudhuri and J.A. Schwartz, {\it Phys. Lett.} {\bf B219}
(1989) 291.
\bibitem{FLI} I. Antoniadis, J. Ellis, J. Hagelin and D.V. Nanopoulos,
{\it Phys. Lett.} {\bf B231} (1989) 65.
\bibitem{FGKP} I. Antoniadis, J. Ellis, E. Floratos, D.V. Nanopoulos
and T. Tomaras, {\it Phys. Lett.} {\bf B191} (1987) 96; S. Ferrara,
L. Girardello, C. Kounnas and M. Porrati, {\it Phys. Lett.} {\bf B192}
(1987) 368, {\bf B194} (1987) 358.
\end{thebibliography}
\end{document}